\documentclass{article}
\usepackage{spconf,amsmath,graphicx}
\usepackage{enumitem}
\usepackage{hyperref}

\def\negvspace{\vspace{-0.8em}}

\title{Overlap-aware diarization:\\Resegmentation using neural end-to-end overlapped speech detection}
\name{Latan\'{e} Bullock $^1$ \qquad Herv\'{e} Bredin  $^2$ \qquad Leibny Paola Garcia-Perera $^3$}
\address{$^{1}$ Rice University, Houston, USA \\
$^{2}$ LIMSI, CNRS, Univ. Paris-Sud, Universit\'{e} Paris-Saclay, Orsay, France \\
$^{3}$ Center for Language and Speech Processing, The Johns Hopkins University, Baltimore, USA \thanks{ The research reported here was conducted at the 2019 Frederick Jelinek Memorial Summer Workshop on Speech and Language Technologies, hosted at L'\'Ecole de Technologie Sup\'erieure (Montreal, Canada) and sponsored by Johns Hopkins University with unrestricted gifts from Amazon, Facebook, Google, and Microsoft. It was also supported by the French National Research Agency (ANR) via the funding of the PLUMCOT project (ANR-16-CE92-0025).}}

\begin{document}
\maketitle

\begin{abstract} %

We address the problem of effectively handling overlapping speech in a diarization system.
First, we detail a neural Long Short-Term Memory-
based architecture for overlap detection.
Secondly, detected overlap regions are exploited in conjunction with a frame-level speaker posterior matrix to make two-speaker assignments for overlapped frames in the  resegmentation step.
The overlap detection module achieves state-of-the-art performance on the AMI, DIHARD, and ETAPE corpora.
We apply overlap-aware resegmentation on AMI, resulting in a 20\% relative DER reduction over the baseline system.
While this approach is by no means an end-all solution to overlap-aware diarization, it reveals promising directions for handling overlap.

\end{abstract}

\section{Introduction}
\label{sec:intro}
Speaker diarization answers the question,  `Who spoke when?' in an audio recording.
In favorable conditions, modern diarization systems are able to achieve error rates nearly on par with those of humans.
However, even the best diarization systems struggle to identify who was speaking in adverse scenarios.
An audio file can be \textit{adverse} in terms of the number of speakers (and, in particular, the amount of overlapping speech), the age of the speakers in the recording, the proximity of the microphone to the speakers, or any combination of these.
There is a need for robust speaker diarization to process child-centered and other naturalistic recordings,
massive amounts of online audio and video, and clinical interviews, to name a just a few.

There have been several studies on overlap detection and its impact on diarization.
One of the first is~\cite{otterson2007efficient}, which investigates how overlap detection could help diarization. %
Later,  the authors in~\cite{boakye2008overlapped} detect overlap with a three-state Hidden Markov Model, and subsequently sum over frame-level posteriors for all of the frames within a segment to make second-speaker assignments. %
\cite{huijbregts2009speech}~propose a `two-pass' system to first detect overlap, then use it to purify speaker models and make assignments.
\cite{yella2012speaker}~used information external to the overlapped speech - namely the surrounding silence - in it’s detection.
\cite{geiger2013detecting}~introduce neural networks to the overlap detection problem. Their main findings are that LSTM-based detection provides comparable results to HMM, and LSTM+HMM is better than HMM. Later in~\cite{andrei2017detecting}, a convolutional neural network (CNN) architecture was used for detection. %
\cite{hagerer2017enhancing} defend the use of artificially mixed data for training in order to combat the imbalance of overlapped and monospeaker regions. %
Most recently,~\cite{Kunesova2019} report CNN-based overlap detection accuracy and evaluate the resulting potential change in diarization error rate (DER), but assume access to perfect two-label assignment.

Some other studies, such as DIHARD I and DIHARD II~\cite{ryant2019second, ryant2018first},
clearly show that handling overlap is crucial and remains an open problem.
In this research, we investigate the use of overlap information to improve diarization performance. Our two-stage process combines detecting overlap in the audio with recurrent neural networks, and hypothesizing two speaker labels in regions with overlap.

\begin{figure*}
    \centering
    \includegraphics[width=0.8\linewidth]{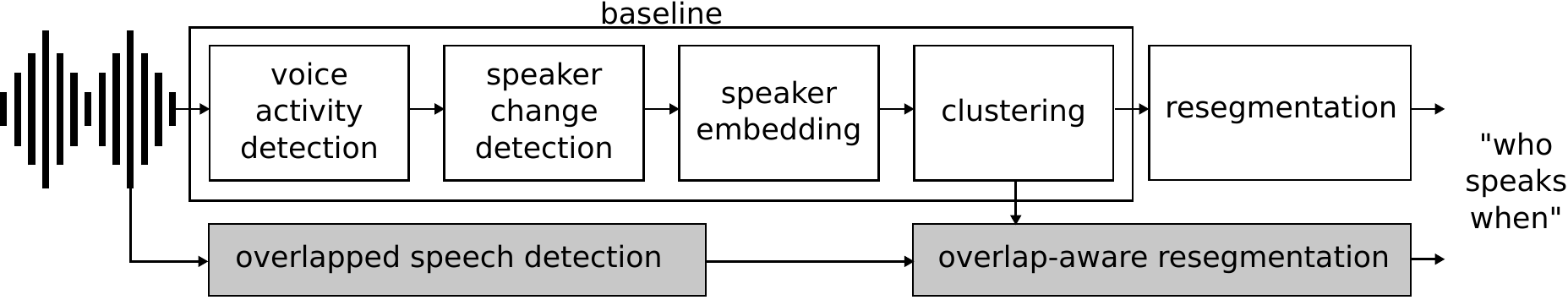}
    \caption{The proposed pipeline for speaker diarization. The baseline incorporates end-to-end neural voice activity detection, speaker change detection, and speaker embeddings, with clustering performed via affinity propagation \cite{Yin2018}. It is available in {\small \texttt{pyannote.audio}} toolkit~\cite{Bredin2020}. The grey boxes highlight the our contributions: neural (LSTM-based) detection of overlapping speech and a simple frame-level resegmentation module designed to account for the overlapping speech. }
    \label{fig:pipeline}
\end{figure*}

\section{Overlapped speech detection}
\label{sec:detection}

Overlapped speech detection is the task of detecting regions where at least two speakers are speaking at the same time.
Detecting regions of overlapped speech is most effectively solved with a temporal approach, where we take into account the sequential nature of speech.

\subsection{Principle}
\negvspace
We address overlapped speech detection as a sequence labeling task where the input is the sequence of feature vectors $\mathbf{X} = \{\mathbf{x}_1, \mathbf{x}_2, \ldots, \mathbf{x}_T\}$ and the expected output is the corresponding sequence of labels $\mathbf{y} = \{y_1, y_2,\ldots, y_T\}$ with $y_{t} = 0$ if there is zero or one speaker at time step $t$ and $y_t = 1$ if there are two speakers or more.

\begin{figure}
    \centering
    \includegraphics{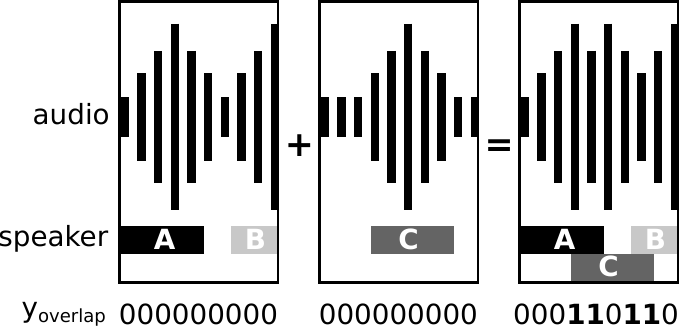}
    \caption{To increase the number of positive training samples for overlapped speech detection, artificial audio chunks are created by summing two random audio chunks.}
    \label{fig:ovl}
\end{figure}

Because processing long audio files of variable lengths is neither practical nor efficient, we rely on shorter fixed-length sub-sequences. At training time, fixed-length sub-sequences are drawn randomly from the training set to form  mini-batches, increasing training sample variability (data augmentation) and training time (shorter sequences). To address the class imbalance problem, half of the training sub-sequences are artificially made from the weighted sum of two random sub-sequences, as depicted in Figure~\ref{fig:ovl}.

At test time, audio files are processed using overlapping sliding windows of the same length as used in training. For each time step $t$, this results in several overlapping sequences of prediction scores, which are averaged to obtain the final score of each class. Finally, time steps with prediction scores greater than a tunable threshold $\theta_{\text{OSD}}$ are marked as overlapped speech.

\subsection{Implementation details}
\negvspace
\begin{figure*}
    \centering
    \includegraphics[width=0.8\linewidth]{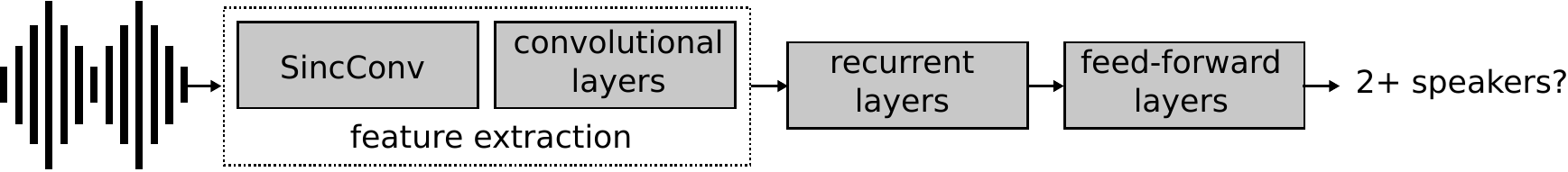}
    \caption{Architecture of the neural network used for end-to-end overlapped speech detection. We also report detection results where the trainable feature extraction part is replaced by handcrafted MFCC features.}
    \label{fig:pyannet}
\end{figure*}

Models are based on the architecture depicted in Figure~\ref{fig:pyannet}.
They are trained on 2s audio chunks, either with handcrafted MFCC features (19 coefficients extracted every 10ms on 25ms windows, with first- and second-order derivatives) or with trainable SincNet features (using the configuration of the original paper~\cite{Ravanelli2018}).
The rest of the network includes two stacked bi-directional Long Short-Term Memory (LSTM) recurrent layers (each with 128 units in both forward and backward directions), two feed-forward layers (128 units, \emph{tanh} activation) and a final classification layer (2 units, \emph{softmax} activation), fed into binary cross-entropy loss.

\section{Overlap-aware resegmentation}
\label{sec:assignment}
While most diarization systems hypothesize a single speaker in all voiced regions, a robust overlapping speech detector opens up the possibility of hypothesizing an additional speaker in overlapping regions.

\subsection{Principle}
\negvspace

\begin{figure*}
    \centering
    \includegraphics[width=0.8\linewidth]{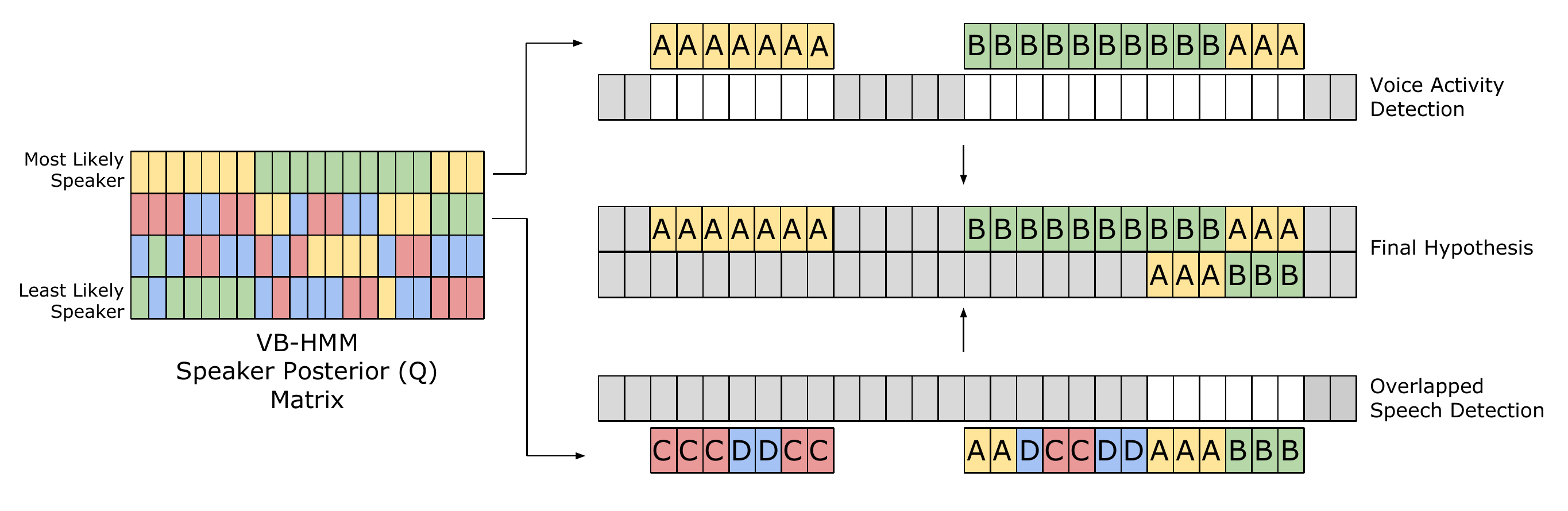}
    \caption{Illustration of proposed method for assigning secondary speakers in overlap regions. The speaker posterior matrix from VB-HMM resegmentation (on the left) serves as the source of speaker label hypotheses. The most likely speaker sequence is masked with the voice activity detection (upper right), while the second most likely speaker sequence is masked by the overlap detection output (lower right). The final diarization hypothesis is the union of the two.}
    \label{fig:assignment}
\end{figure*}

Depicted in Figure~\ref{fig:assignment}, our proposed approach relies heavily on the i-vector-based Variational Bayes Hidden Markov Model (VB-HMM) introduced for speaker diarization in~\cite{diez2018speaker}, and applied to resegmentation in~\cite{sell2015diarization}.
We use the output of the speaker diarization baseline as the binary initialization of the per-frame speaker posterior matrix: $\boldsymbol{Q}_{st}$ is initialized to $1$ if speaker $s$ is responsible for the speech at the voiced frame $t$, and $0$ otherwise. After VB-HMM resegmentation, the previously one-hot hard assignments of speakers to frames in $\boldsymbol{Q}$ become soft probabilities.
The most likely speaker is assigned to frames detected as speech by the voice activity detector.
A second most likely speaker is only assigned for frames detected as overlapped speech.

\subsection{Implementation details}
\negvspace

We first perform resegmentation using \cite{diez2018speaker}'s VB-HMM module.
Feature vectors for the module are length-60 MFCCs with deltas and double deltas, extracted in 10ms steps with a 25ms window.
The (400-dimensional) i-vector extractor and diagonal (1024-component) Universal Background Model are trained on the training portion of AMI Headset mix.
We use a single VB inference iteration (default 10)
and adjust the loop probability parameter to 0.95 (default 0.9). Otherwise, we keep the default parameters of the VB-HMM diarization module.

\section{Experiments}

Overlapped speech detection models were trained, tuned, and tested on three different datasets whose statistics are summarized in Table~\ref{tab:stats}:
\begin{itemize}[noitemsep]
\item AMI (Headset  mix)~\cite{ami} is a subset of the AMI corpus that consists of summed recordings of spontaneous speech of mainly four speakers; %
\item DIHARD II~\cite{ryant2019second} contains single channel wide-band audio from 11 challenging domains that range from very clean (near-field recordings of read audiobooks) to
noisy, far-field recordings; %
\item ETAPE (TV subset)~\cite{etape} consists of TV content in French (news, talk shows, debates). %
\end{itemize}

\begin{table*}[htb]
    \centering
    \begin{tabular}{|lc|ccc|ccc|ccc|}
        \hline
        \textbf{Dataset} & & \multicolumn{3}{c}{\textbf{Train}} & \multicolumn{3}{c}{\textbf{Development}} & \multicolumn{3}{c|}{\textbf{Evaluation}} \\
        \hline
        AMI (Headset mix) & \cite{ami} & 70h & {\small 85\%} & {\footnotesize 19\%} & 14h & {\small 84\%} & {\footnotesize 20\%} & \textbf{14h} & \textbf{{\small 82\%}} & \textbf{{\footnotesize 19\%}}\\
        DIHARD II & \cite{dihard} & 15h & {\small 75\%} & {\footnotesize 9\%} & 8h & {\small 77\%} & {\footnotesize 11\%} & 22h & {\small 74\%} & {\footnotesize 9\%}\\
        ETAPE (TV) & \cite{etape} & 14h & {\small 94\%} & {\footnotesize 6\%} & 4h & {\small 93\%} & {\footnotesize 5\%} & 4h & {\small 92\%} & {\footnotesize 7\%}\\
        \hline
    \end{tabular}
    \caption{Datasets statistics. For each subset, we report the total audio duration (in hours), the amount of speech (as percentage of audio duration), and the amount of overlapped speech (as percentage of speech duration). For instance, AMI evaluation set amounts to 14h of audio, 82\% of which is speech (11.5h), among which 19\% is overlapped speech (2.2h). Note that DIHARD does not come with a training set so the official development set was divided into two thirds for training and one third for development. }
    \label{tab:stats}
\end{table*}

The proposed overlap-aware resegmentation module has only been tested on AMI Headset mix. 81\% of the total speech in voiced regions is single-speaker and 15\% of the time two-speaker, leaving approximately 4\% of the time to three or more speakers. This implies that the two-speaker situation accounts for about 75\% of the overlap regions -- justifying our initial focus on this case.

Code, configuration files, and pre-trained models for reproducing the speaker diarization baseline and overlapped speech detection results are available in the {\small \texttt{pyannote.audio}} repository~\cite{Bredin2020}. Code for VB-HMM resegmentation is provided by Brno University of Technology\footnote{\url{https://speech.fit.vutbr.cz/software}}, and all assignment code can be found in the JSALT 2019 Speaker Detection team repository\footnote{\url{https://github.com/jsalt2019-diadet/jsalt2019-diadet}}.
The {\small \texttt{pyannote.metrics}} toolkit~\cite{pyannote.metrics} is used to evaluate overlapped speech detection in terms of precision and recall, and resegmentation in terms of diarization error rate (DER). DER is the portion of the recording that is labelled incorrectly, with three possible types of errors: false alarm, missed detection, and speaker confusion.

\subsection{Overlapped speech detection}
\negvspace
As reported in Table~\ref{tab:ovl}, the end-to-end variant consistently outperforms the one based on handcrafted features for all datasets, setting a new state-of-the-art performance on all three datasets\footnote{Thanks to Claude Barras for providing the overlapped speech detection output corresponding to system $L_1$ in Table 2 of~\cite{Charlet2013}, and Marie Kune{\v{s}}ov{\'a} for providing the overlapped speech detection output corresponding to system \emph{"AMI test (all subsets) + dereverberation"} in Table 2 of~\cite{Kunesova2019}.} -- though we could not find any previously published overlapped speech detection results for DIHARD.
When tuned for high (90\%) precision, the proposed approach gets a very low recall of 1.5\% on DIHARD, making it almost useless for the subsequent overlap assignment step.

\begin{table*}[htb]
    \centering
    \begin{tabular}{|l|l|l|l|l|l|l|}
        \hline
        & \multicolumn{2}{c|}{\textbf{AMI}} & \multicolumn{2}{c|}{\textbf{DIHARD}} & \multicolumn{2}{c|}{\textbf{ETAPE}}\\
        & Precision & Recall & Precision & Recall & Precision & Recall \\
        \hline
        Baseline                    & 75.8 {\scriptsize{80.5}} \cite{Kunesova2019} & 44.6 {\scriptsize{50.2}} \cite{Kunesova2019} & & & 60.3 \cite{Charlet2013} &  52.7 \cite{Charlet2013} \\
        \hline
        Proposed (MFCC) &  91.9 \scriptsize{90.0} & 48.4 \scriptsize{52.5} & 58.0 \scriptsize{73.8} & 17.6 \scriptsize{14.0} & 67.1 \scriptsize{55.0} & 57.3 \scriptsize{55.3} \\
        Proposed (waveform) & 86.8 \scriptsize{90.0} & 65.8 \scriptsize{63.8} & 64.5 \scriptsize{75.3} & 26.7 \scriptsize{24.4} & 69.6 \scriptsize{60.0} & 61.7 \scriptsize{63.6} \\
        \hline

    \end{tabular}
    \caption{Evaluation of overlapped speech detection models, in terms of precision (\%) and recall (\%). Results on the development set are reported using small font size. We report two variants: the first one is based on handcrafted features (MFCCs) and the other one is an end-to-end model processing the waveform directly. \emph{Baseline} corresponds to the best result we could find in the literature as of October 2019.}
    \label{tab:ovl}
\end{table*}

\subsection{Overlap-aware resegmentation}
\negvspace
 \begin{table}[]
    \centering
    \begin{tabular}{|l|c|c|c|c|}
        \hline
                        & DER\% &    FA\% & Miss\% &    Conf\% \\
        \hline
        Baseline    &    29.7 &    \textbf{3.0} &    20.8 &    5.8 \\
        {\small + VB resegmentation}         &    28.9 &    \textbf{3.0} &    20.9 &    \textbf{5.0} \\
        {\small + overlap assignment}         &    \textbf{23.8} &    3.6 &    \textbf{13.0} &    7.2 \\
        \hline
        {\small + oracle detection}         &    22.2 &    3.1 &    6.0 &    13.2 \\
        \hline
        {\small + oracle assignment} & 11.8 & 0.6 &  11.2 &   0.0 \\
        \hline
    \end{tabular}
    \caption{AMI Headset mix diarization, false alarm, missed detection, and speaker confusion error rates after VB-HMM resegmentation and overlap assignment. The proposed assignment technique using oracle overlap detection and using oracle assignment are also reported. Oracle assignment refers to ideal both primary and secondary speaker labels. }
    \label{tab:ovl-assignment}
\end{table}

The impact of our second contribution on the performance of the diarization pipeline is reported in Table~\ref{tab:ovl-assignment}. Overall, our proposed overlap-aware resegmentation approach brings a significant 20\% relative (or 5.9\% absolute) improvement in terms of diarization error rate (from 29.7\% down to 23.8\%).

A detailed analysis shows that the VB-HMM resegmentation step reduces confusion error by less than 1\% while leaving -- by design -- false alarm and miss detection rates unchanged. Tuned for high precision, overlapped speech detection reduces missed detection by 38\% relative (or 7.9\% absolute), at the expense of a small increase in false alarm rate (from 3.0\% to 3.6\%). The secondary speaker assignment does increase speaker confusion by more than 2\% (out of the 7.9\% of correctly detected overlapped speech regions). Overall, the combination of our two contributions (overlapped speech detection and assignment) leads to a new state of the art on AMI Headset Mix, by a large margin.

Switching to oracle overlapped speech detection only brings a minor performance boost (from DER=23.8\% down to 22.2\%). This indicates that most future improvements will likely come from a better speaker assignment -- which is confirmed by the oracle assignment experiment.

-%

\section{Conclusion}
\label{sec:conclusion}

In this paper we have highlighted two contributions to the speaker diarization task. The first is an neural architecture for overlap detection, and the second is a simple yet effective resegmentation module that assigns two speakers in frames detected as overlapping speech.

The overlap detector predicts overlapping speech regions with state-of-the-art accuracy on AMI and ETAPE, and sets the baseline for future experimentation on DIHARD II.
Our proposed solution for overlap-aware resegmentation was tested on AMI and beats state-of-the-art systems in DER due to a drastic decrease in missed detection error. However, further work is needed to more accurately assign secondary speakers, as evidenced by the large increase in speaker confusion error with oracle detection. Additionally, testing on other datasets will be necessary to establish the method as a robust approach.

Our hope is that the present study will encourage more research on both overlap detection and its practical uses in diarization systems. Two of the primary remaining questions are: how to increase accuracy of secondary speaker assignment, and how to extend assignment to more than two speakers. A system inspired by~\cite{ding2019personal} could integrate the detection and assignment steps to improve secondary speakers hypotheses.
\cite{Stter2019CountNetET} recently proposed a neural architecture to count the number of concurrent speakers in a signal, which could enable three or more speaker assignment.

\bibliographystyle{IEEEbib}
\bibliography{refs}

\end{document}